%% file: main.tex
\documentclass{article}
\usepackage{spconf}
\usepackage{cite}
\usepackage{amsmath,amssymb,amsfonts,MnSymbol}
\usepackage{algorithmic}
\usepackage{graphicx}
\usepackage{textcomp}
\usepackage{xcolor}
\usepackage{balance}
\usepackage{acronym}
\usepackage{hyperref}
\usepackage{csquotes}
\usepackage{adjustbox}
\usepackage{nicematrix}
\usepackage{booktabs}
\usepackage{multirow}
\usepackage[group-separator={,}]{siunitx}

\makeatletter
\newcommand*{\org@overidelabel}{}
\let\org@overridelabel\AC@verridelabel
\renewcommand*{\AC@verridelabel}[1]{%
  \@bsphack
  \protected@write\@auxout{}{\string\AC@undonewlabel{#1@cref}}%
  \org@overridelabel{#1}%
  \@esphack
}%
\makeatother

\usepackage{cleveref}
\usepackage{tikz,pgfplots}
\pgfplotsset{compat=1.18}
\usepgfplotslibrary{colorbrewer}

\usepackage{threeparttable}

\newcommand{\BE}{\begin{equation*}\begin{aligned}}
\newcommand{\EE}{\end{aligned}\end{equation*}}

\DeclareMathOperator{\avgpool}{AvgPool}
\DeclareMathOperator{\bn}{BN}
\DeclareMathOperator{\relu}{ReLU}

\acrodef{aqa}[AQA]{audio question-answering}
\acrodef{ast}[AST]{audio spectrogram transformer}
\acrodef{cav-mae}[CAV-MAE]{contrastive audio-visual masked auto-encoder}
\acrodef{ipd}[IPD]{interaural phase difference}
\acrodef{ild}[ILD]{interaural level difference}
\acrodef{stft}[STFT]{short-time Fourier transform}
\acrodef{gcc}[GCC-PHAT]{Generalized Cross-Correlation with Phase Transform}
\acrodef{dspast}[DSpAST]{disentangled Spatial\ac{ast}}
\acrodef{llama}[LLaMA]{large language model Meta AI}
\acrodef{llm}[LLM]{large language model}
\acrodef{lora}[LoRA]{low-rank adaptation}
\acrodef{ltu}[LTU]{listen, think and understand}
\acrodef{qlora}[qLoRA]{quantized low-rank adaptation}
\acrodef{cot}[CoT]{chain-of-thought}
\acrodef{map}[mAP]{mean average precision}
\acrodef{doae}[DoAE]{direction-of-arrival estimation}
\acrodef{dp}[DP]{distance prediction}
\acrodef{der}[DER]{distance error rate}
\acrodef{acc}[Acc]{accuracy}
\acrodef{ba}[BAcc]{binary accuracy}
\acrodef{sed}[SED]{sound event detection}
\acrodef{mae}[MAE]{mean angular error}
\acrodef{er20}[$\mathop{\text{ER}}_{\ang{20}}$]{error rate at \ang{20}}
\acrodef{rir}[RIR]{room impulse response}

\begin{document}
\ninept
\title{DSpAST: Disentangled Representations for Spatial Audio Reasoning with Large Language Models}

\name{Kevin Wilkinghoff$~^{1,2}$, Zheng-Hua Tan$^{1,2}$}
\address{$^{1}$Department of Electronic Systems, Aalborg University, Denmark, $^{2}$Pioneer Centre for AI, Denmark}

\maketitle

\begin{abstract}
Reasoning about spatial audio with large language models requires a spatial audio encoder as an acoustic front-end to obtain audio embeddings for further processing.
Such an encoder needs to capture all information required to detect the type of sound events, as well as the direction and distance of their corresponding sources.
Accomplishing this with a single audio encoder is demanding as the information required for each of these tasks is mostly independent of each other.
As a result, the performance obtained with a single encoder is often worse than when using task-specific audio encoders.
In this work, we present \acs{dspast}, a novel audio encoder based on Spatial\acs{ast} that learns disentangled representations of spatial audio while having only $0.2\%$ additional parameters.
Experiments on SpatialSoundQA with the spatial audio reasoning system BAT demonstrate that \acs{dspast} significantly outperforms Spatial\acs{ast}.
\end{abstract}

\begin{keywords}
audio reasoning, spatial audio, large language models, disentanglement, sound source localization
\end{keywords}

\acresetall

\section{Introduction}
Humans and numerous other animals use two auditory sensors, the so-called ears, to collect spatial acoustic information \cite{koyama2025past}, allowing them not only to distinguish between different sound events, but also to localize sound sources by estimating their distance and direction to make sense of the world.
Since evolution favored this binaural information over monaural hearing, intelligent systems that perceive the world through audio should also use multiple microphones.
Only then can these systems use spatial audio information to separate and detect sound events in complex acoustic scenes \cite{yasuda2025description,nguyen2025baseline} and localize sound sources, such as, for example, animals \cite{peterson2024vocal}.
\par
Multimodal \acp{llm} that utilize additional data modalities in addition to textual data have emerged as a powerful tool for many applications.
In the context of audio, an example is audio reasoning systems whose task is to answer questions about provided audio recordings \cite{gong2024listen,tang2024can,zheng2024bat,yang2025multi,liang2025acoustic}.
Usually, training such a system is achieved by mapping audio signals and questions with designated encoders into a shared embedding space that can be understood by a pre-trained \ac{llm} and then fine-tuning the \ac{llm}.
Most general-purpose audio encoders, such as \ac{ast} \cite{gong2021ast} or spatial audio-based variants such as SpatialAST \cite{zheng2024bat} are pre-trained on Audioset \cite{gemmeke2017audioset}, which is the largest publicly available audio dataset.
There is also some evidence that well-performing audio encoders of speech recognition systems may be suitable for general audio tagging \cite{gong2023whisper}.
\par
The caveat of training a spatial audio encoder is that they have to solve multiple tasks, namely \ac{sed}, \ac{dp} and \ac{doae}, and often improving the performance of one task negatively affects the performance of the other tasks \cite{zheng2024bat}.
Intuitively, this makes sense as the sound event, the distance and direction of a sound source are mostly independent of each other, e.g. the distance of a sound source can be the same for arbitrary sound events and vice versa, yet the representations used for determining these attributes are not.
\par
The goal of this work is to present an audio encoder architecture that is able to effectively handle all three tasks at once.
To this end, we propose to disentangle the representations into information needed for each specific task, similar to disentangling speech representations into phonetic, prosodic, and speaker information \cite{xie2023improved,xie2024speaker,aihara2025exploring,khurana2025factorized}.
A naive solution to disentangling the information about sound event types, their distance, and direction is to use task-specific audio encoders.
However, this increases the number of parameters by a factor of three.
An efficient solution provided by a single model requires a more sophisticated strategy.
\par
In this work, we present \ac{dspast}, which is a disentangled spatial audio encoder utilizing multiple audio features that are combined with feature attention modules in task-specific branches.
This allows the model to choose the most useful audio features for each specific task without negatively impacting the performance for other tasks.
As a result, we achieve significantly higher performance than the baseline model, SpatialAST, with approximately the same model size.
The source code and checkpoints of our system are publicly available\footnote{\url{https://github.com/wilkinghoff/DSpAST/}}.

\section{Review of SpatialAST}
\label{sec:spatialast}
SpatialAST, the spatial audio encoder of the audio reasoning system BAT \cite{zheng2024bat}, which is based on Llama-2 7B \cite{zhang2024llama}, serves as a baseline model in this work.
The goal of SpatialAST is to extract representations of binaural audio suitable for \ac{sed}, \ac{doae} and \ac{dp}.
After training, the frozen audio encoder is used to generate embeddings from spatial audio and then a Q-Former \cite{zhang2024vision} maps the audio embeddings into the text-embedding space.
Finally, BAT is fine-tuned on pairs of binaural audio clips and questions/answers belonging to different reasoning tasks.
In the following, we briefly review the front-end, backbone and training objective of SpatialAST.
\par
\textbf{Front-end:}
SpatialAST uses log-mel spectrograms of the left and right channels and their \acp{ipd} as input features.
The log-mel spectrograms are based on $M=128$ mel bins and a \ac{stft} with a Hanning window of size $1024$ and a hop size of $320$.
Furthermore, a bandpass filter with a passband between $\SI{50}{\hertz}$ and $\SI{14000}{\hertz}$ is applied.
For the \acp{ipd}, phase spectrograms are computed first.
Then, the cosine and sine transformation and the same mel filterbank as used for the log-mel spectrograms are applied such that all features share the same dimensions.
As a final step of the front-end, all four features are stacked into a single tensor for further processing.
\par
\textbf{Backbone:}
\label{sec:spatialast_backbone}
The backbone consists of three modules: a feature fusion module, a patch embedding module, and a transformer module.
The feature fusion module consists of a two-dimensional convolution with a kernel size of $3\times3$ and effectively takes the element-wise arithmetic mean over all features after the convolution, followed by batch normalization \cite{ioffe2015batch} and a Gaussian error linear unit \cite{hendrycks2016bridging}.
From this, patch embeddings are generated using a two-dimensional CNN with a kernel size and stride of $16\times16$ to extract non-overlapping patches.
Furthermore, three CLS tokens are appended, one for each of the three tasks \ac{sed}, \ac{dp} and \ac{doae}.
Lastly, a transformer consisting of twelve encoder blocks is utilized to process the patch embeddings.
\par
\textbf{Training objective:}
SpatialAST is trained to predict all types of sound events that occur and the distance, azimuth, and elevation of the sound sources.
The distances and angles are discretized into intervals of $\SI{0.5}{\meter}$ and $\SI{1}{\degree}$, respectively.
To this end, linear classification heads are used.
For the \ac{sed} task, a sigmoid activation function with a binary cross-entropy is used to allow the detection of multiple sound events.
For each of the other attributes, a softmax activation function with a categorical cross-entropy is used.
The total loss is a weighted sum of the task-specific losses $\mathcal{L}_\text{\ac{sed}}, \mathcal{L}_\text{\ac{dp}}, \mathcal{L}_\text{\ac{doae}}$ given by
\BE \mathcal{L}_\text{total}=\lambda_1\mathcal{L}_\text{\ac{sed}}+\lambda_2\mathcal{L}_\text{\ac{dp}}+\lambda_3\mathcal{L}_\text{\ac{doae}}\EE
where the weights $\lambda_1,\lambda_2,\lambda_3\in\mathbb{R}_+$ are tunable hyperparameters and $\mathcal{L}_\text{\ac{doae}}$ is the sum of losses for azimuth and elevation.

\section{DSpAST: Disentangled Spatial Audio Encoder}
\input{figures/overview}
As the name suggests, \ac{dspast} is an extension of SpatialAST that learns disentangled representations for the three tasks, \ac{sed}, \ac{dp} and \ac{doae}, required for spatial audio reasoning.
To this end, we made three modifications to the original architecture by introducing additional spatial audio features, a feature attention module, and task-specific branches.
In the following, we will describe all three modifications.
An illustration of \ac{dspast} can be found in \Cref{fig:overview}.

\subsection{Additional spatial audio features}
In addition to the audio features used by SpatialAST, other commonly used features for joint sound event detection and localization from binaural audio include the mean mel spectrograms of the left and right channels, \ac{ild}, and \ac{gcc} \cite{krause2022binaural,krause2024binaural}.
Since our proposed encoder employs a feature attention module (cf.~\Cref{sec:feature_attention}) that can select the most suitable features for each task, effectively ignoring features that are not beneficial, we also include \ac{ild} and \ac{gcc}, as implemented in \cite{cao2019polyphonic}.
Note that the mean of the log-mel spectrograms from the left and right channels does not need to be included explicitly because the weighted sum of the input features used by \ac{dspast} already accounts for this.

\subsection{Feature attention module}
\label{sec:feature_attention}
Different spatial audio features are of different importance for identifying sound event classes, predicting distance, or estimating direction-of-arrival.
Therefore, it is favorable if the model can select the most informative features for each specific task.
As stated in \Cref{sec:spatialast}, Spatial\acs{ast} takes a weighted mean of all features using a CNN.
This has the advantage that the model can handle an arbitrary number of features using the same architecture but uses the same feature weights for each task.
Similarly to a channel attention mechanism \cite{woo2018cbam}, we propose a feature attention module to allow for a task-specific weighting of the features before taking their mean.
In addition to improving the resulting performance, this also explains the results by highlighting which spatial features are most informative for each task (see \Cref{sec:attention}).
\par
More concretely, we used the following feature attention module.
Let $B,C,T,M\in\mathbb{N}$ denote the batch size, the number of features, the temporal dimension, and the number of mel bins, respectively.
First, feature attention masks $\mathcal{M}(x)\in\mathbb{R}^{B\times C\times 1\times 1}$ for a batch of input samples $x\in\mathbb{R}^{B\times C\times T\times M}$ are computed as
\BE \mathcal{M}(x):=\sigma(W_1\relu(W_0\bn(\avgpool(x))+b_0)+b_1)\EE
where $\sigma$ denotes the sigmoid activation function, $\relu$ denotes the ReLU activation function, $\bn$ refers to a batch normalization layer \cite{ioffe2015batch}, $\avgpool$ represents a global average pooling operation, and $W_0,W_1,b_0,b_1$ denote the weights and bias terms of two dense layers.
These attention masks are applied to the input feature tensors with element-wise multiplication.

\subsection{Disentangled representations}
To enable \ac{dspast} to incorporate task-specific feature attention modules, it is required to separate the information flow for each task.
This is achieved by introducing three task-specific branches, each influenced solely by the loss associated with its corresponding task.
Each branch consists of a feature attention module, the three backbone modules of SpatialAST (cf.~\Cref{sec:spatialast_backbone}), and a linear projection that reduces the representation dimensionality to one-third of its original size.
This ensures that the projection head size does not increase when fine-tuning an \ac{llm} for the spatial audio reasoning task.
After that, the resulting representations from each branch are stacked, resulting in CLS and audio representations of the same dimension as in SpatialAST.
In contrast to the representations of SpatialAST, however, this procedure effectively disentangles audio and class representations into three task-specific parts for \ac{sed}, \ac{dp} and \ac{doae}.
For training \ac{dspast}, the mean of the three task-specific CLS tokens is used as input to the corresponding classification head, and the same loss as for SpatialAST is used.
An important aspect of the design is that the parameters of the transformer and patch embedding modules are shared across three tasks.
This ensures that the model size of \ac{dspast} is almost the same as the size of SpatialAST.
When comparing the number of parameters, SpatialAST has $85.96$ million parameters and \ac{dspast} $86.09$ million parameters, i.e. less than $0.2\%$ additional parameters are used.

\section{Experimental Evaluations}
For the experimental evaluations in this work, we closely followed the experimental setup described in \cite{zheng2024bat}.
Since retraining and evaluating the considered models is computationally expensive, and confidence intervals based on a normal approximation mainly depend on the test dataset size, which is sufficiently large in this case, we report only point estimates of the performance metrics. 

\subsection{Performance of \acs{dspast}}
\input{tables/training_curriculum}
As a first experiment, we compare \ac{dspast} with SpatialAST \cite{zheng2024bat}, conduct ablation studies, and analyze the feature attention weights.

\subsubsection{Experimental setup}
\textbf{Dataset:} For all evaluations, we used the binaural audio dataset described in \cite{zheng2024bat}, which consists of simulated three-dimensional acoustic scenes generated by convolving audio clips from Audioset \cite{gemmeke2017audioset} with \acp{rir} from SoundSpaces 2.0 \cite{chen2022soundspaces} and then padding or trimming them to a duration of $\SI{10}{\second}$.
Before convolution, the loudness of all audio clips is normalized by rescaling the energy of each signal to one.
As some of the Audioset clips contain strong noise, all clips with labels indicating noise were removed from the dataset, resulting in $355$ remaining classes with a total of $1,861,750$ and $18,373$ clips in the unbalanced and balanced split, respectively, and $17,148$ clips in the evaluation set.
\par
\textbf{Evaluation metrics:}
To evaluate the performance of the audio encoders in the three tasks, four different performance metrics were used.
We used the threshold-independent metrics \ac{map} and \ac{mae}, which measures the absolute difference in angle, for the \ac{sed} and \ac{doae} tasks, respectively.
Furthermore, two threshold-based metrics that allow for small deviations from the ground truth of spatial location were used \cite{mesaros2019joint}: 
\ac{er20}, which allows errors up to $\SI{20}{\degree}$, and \ac{der}, which allows errors up to $\SI{0.5}{\meter}$.
\par
\textbf{Implementation details:}
For our implementation of \ac{dspast}, we used the same parameter settings as for the front-end of SpatialAST resulting in features with a time dimension of $T=1024$ and a number of mel bins $M=128$.
Furthermore, the weights of \ac{dspast} prior to training stage 1 are initialized with the official checkpoint of Audio\acs{mae} \cite{huang2022masked} and patch masking with a ratio of $0.25$ for time and frequency is applied during training.
The main differences from SpatialAST are the modified training curriculum shown in \Cref{tab:curriculum}, which consists of three stages instead of two, and using the AdaCos loss \cite{zhang2019adacos} for the \ac{dp} and \ac{doae} task.

\subsubsection{Experimental results}
\input{tables/ast_performance}
\input{tables/bat_performance}
The results can be found in \Cref{tab:ast-performance} and the following observations can be made.
First, \ac{dspast} (stage 3) significantly outperforms SpatialAST for all performance metrics, showing the benefits of our proposed architecture.
Second, although \ac{dspast} already has better performance than SpatialAST after training stage 2, adding stage 3 further improves the \ac{sed} performance while not negatively impacting the performance for \ac{dp} and \ac{doae}.
Furthermore, the added spatial features substantially improve the \ac{doae} performance while yielding a similar performance for the \ac{sed} task and degrading the performance for the \ac{dp} task.
However, we want to note that the \ac{dp} performance is relatively noisy with strong deviations of a few percentage points during training.
Another observation is that feature attention improves performance for all tasks except for the \ac{doae} task, where the performance is very similar regardless of whether or not feature attention is used.

\subsubsection{Attention weight analysis}
\label{sec:attention}
\input{figures/att_wts}
The average feature attention weights for each branch of \ac{dspast} are depicted in \Cref{fig:att_wts}.
Although all branches utilize log-mel spectrograms, the main observation is that the \ac{sed} branch relies on less information provided by the other features compared to the other branches.
This is most apparent for the \ac{gcc} feature, which is almost completely ignored by the feature attention module of the \ac{sed} branch but plays an important role for the \ac{dp} and \ac{doae} task.

\subsection{Performance of \acs{dspast} within BAT}
As a second experiment, we test whether the performance gains of \ac{dspast} on the binaural dataset transfer to the audio reasoning task using its spatial audio representations as input for BAT \cite{zheng2024bat}.

\subsubsection{Experimental setup}
\textbf{Dataset:}
All experiments were carried out using SpatialSoundQA \cite{zheng2024bat}, which consists of pairs of binaural audio clips and questions/answers of five different types labeled A to E.
Audio clips belonging to type A and C have one sound source, and samples of type B, D, and E have two sources.
All questions of type A and B are \ac{sed} tasks and require listing all sound events contained in the clip.
The questions of type B and D are joint \ac{doae} and \ac{dp} tasks and require to provide directions using \enquote{left}/\enquote{right}, \enquote{front}/\enquote{behind}, and \enquote{above}/\enquote{below} as well as distances in $\SI{0.5}{\meter}$ increments.
Answering binary questions of type E with \enquote{yes} or \enquote{no} requires separating the sound sources based on the sound events contained in the audio clip, localizing the sound sources and analyzing their relationship.
For further details, the reader is referred to \cite{zheng2024bat}.
\par
\textbf{Evaluation metrics:}
To measure \ac{sed} performance, \ac{map} is approximated by first computing sentence embeddings for the predictions and labels using \emph{MiniLM-L6-v2} \cite{wang2020minilm} from sentence-transformers \cite{reimers2019sentence-bert}, followed by computing the mean cosine similarity between these embeddings over the test set.
Note that the same strategy is used in \cite{gong2024listen} and \cite{zheng2024bat}, albeit using other sentence embeddings.
\Ac{dp} performance is measured by determining whether the predicted distance is within $\SI{0.5}{\meter}$ of the ground truth and computing a \ac{der} based on that.
For \ac{doae}, all combinations of provided directions \enquote{left}/\enquote{right}, \enquote{front}/\enquote{behind}, and \enquote{above}/\enquote{below} are viewed as eight different classes and accuracy is used.
Similarly, binary accuracy is used for the possible answers \enquote{yes}/\enquote{no} of the reasoning tasks.
\par
\textbf{Implementation details:}
In our experiments, we used the representations of either SpatialAST or \ac{dspast} as input to BAT and fine-tuned with \ac{lora} \cite{hu2022lora}.
Multi-stage training of BAT consists of three stages:
In stage 1, only samples of type A and B are used, in stage 2 samples of type C and D are added, and in stage 3 the entire dataset is used.
Single-stage training corresponds to only training in stage 3. 
In each stage, BAT is trained for five epochs, including one warm-up epoch, using a base learning rate $\num{1e-4}$ and a batch size $4\times 8$.
The model receives binaural audio and questions and is trained to autoregressively predict the next answer token.
During inference, greedy decoding is used.

\subsubsection{Experimental results}
The results can be found in \Cref{tab:bat-performance}.
Two observations can be made.
First, using \ac{dspast} instead of SpatialAST as a tokenizer leads to better performance for all considered tasks.
Second, multi-stage training of BAT based on DSpAST does not offer any performance improvements.
Therefore, we propose using single-stage training to reduce computational costs and maximize the resulting performance.

\section{Limitations and Future Work}
Despite significant performance gains, the performance obtained with \ac{dspast} remains imperfect, and further improvements are needed.
One possible approach is is to incorporate other spatial features, for example, replacing \ac{gcc} with neural GCC features \cite{salvati2021time,berg2022extending}.
Although binaural recordings can work well for sound source localization \cite{wilkins2023two}, using more microphones is likely to enable much more powerful localization capabilities.
Additional gains may be achieved by taking the geometry of the microphone array into account \cite{heikkinen2025gen-a}.
Another promising direction is to improve modeling of sound propagation by generating \acfp{rir} for simulating different rooms \cite{roman2024spatial} or using neural acoustic fields \cite{ick2024spatially}.
\par
Apart from improving performance, spatial audio reasoning systems are not yet on par with human capabilities, and several task extensions can be considered.
Stationary microphones can be replaced with moving ones \cite{krause2024binaural}, and reasoning systems should also perform robustly in noisy environments, for example, by distilling noise-robust spatial audio features \cite{bovbjerg2025learning}.
Combining spatial audio with other modalities such as visual data \cite{chen2025savvy} is another direction.
Finally, moving from an audio tagging task to true \ac{sed}, which requires temporal localization of sound events, and enabling open reasoning, as in \acs{ltu} \cite{gong2024listen}, are key steps toward spatial audio intelligence.

\section{Conclusion}
In this work, we presented \ac{dspast}, a disentangled spatial audio encoder consisting of task-specific branches for \ac{sed}, \ac{dp} and \ac{doae}.
Each branch utilizes log-mel spectrograms, \ac{ipd}, \ac{ild}, and \ac{gcc} as input to a feature attention module.
In experiments with binaural audio clips, \ac{dspast} significantly outperformed the baseline SpatialAST while only having $0.2\%$ additional parameters.
In addition, attention weights were shown to help interpret the results, and improved performance was observed when using \ac{dspast} as a tokenizer in a spatial audio reasoning model evaluated on SPATIALSOUNDQA.

\section{Acknowledgments}
The authors thank Nikolai Lund K\"{u}hne for technical discussions.

\clearpage
\bibliographystyle{IEEEbib-abbrev}
\bibliography{refs}

\end{document}

%% file: figures/overview.tex
\begin{figure}[t]
    \vspace{-6pt}
    \centering
    \begin{adjustbox}{max width=\columnwidth}
          \includegraphics{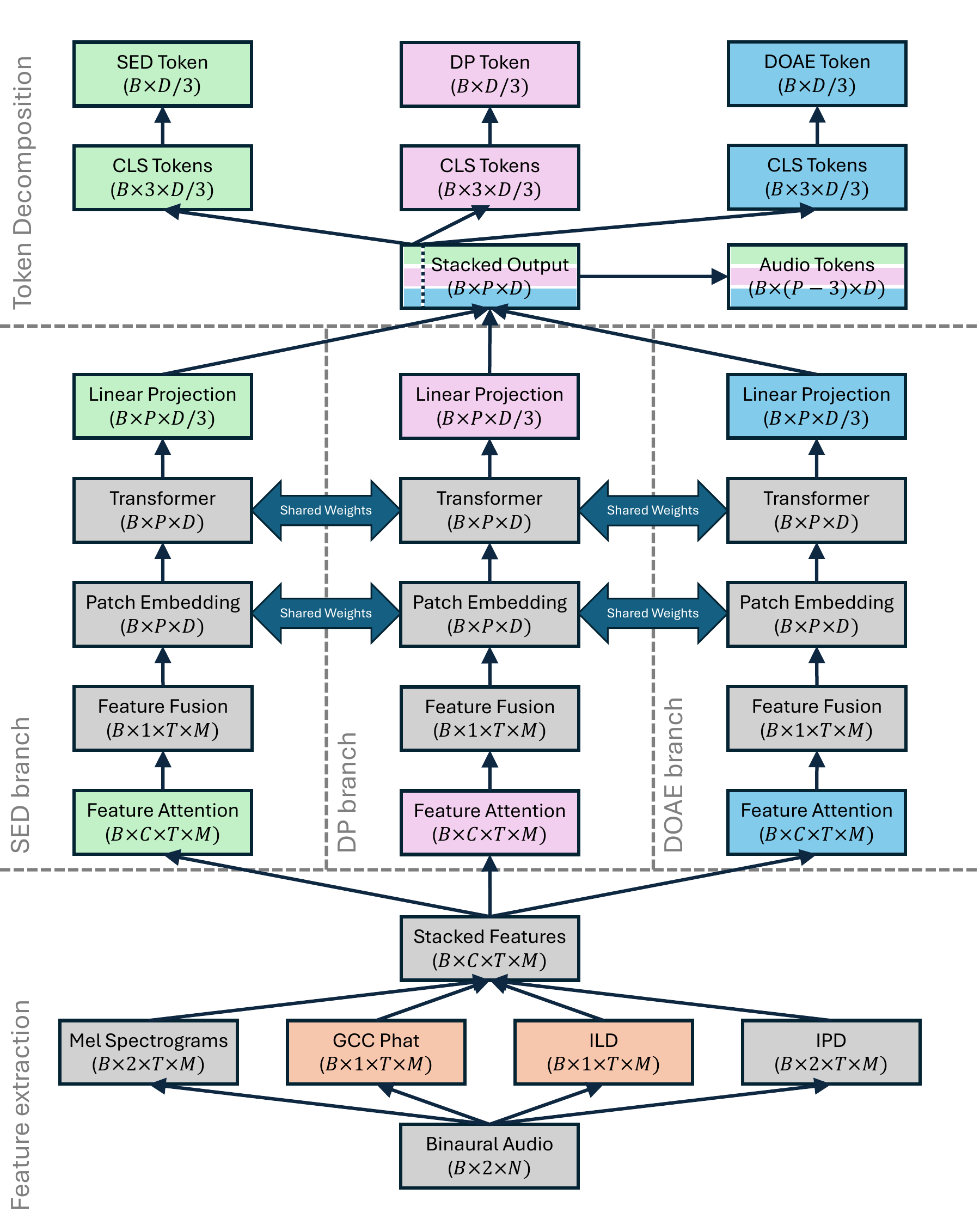}
    \end{adjustbox}
    \caption{Architecture of \acs{dspast}. Parts in gray correspond to the architecture of Spatial\ac{ast} \cite{zheng2024bat}, colored parts are our modifications.}
    \label{fig:overview}
\end{figure}

%% file: tables/training_curriculum.tex
\begin{table}
    \centering
    \vspace{-6pt}
    \caption{Training curriculum of \acs{dspast}.}
    \begin{adjustbox}{max width=\columnwidth}
    \begin{NiceTabular}{l ccc}
        \toprule
         & stage 1 & stage 2 & stage 3\\
        \midrule
        Audioset split&unbalanced&unbalanced&balanced\\
        weighted sampling&yes&yes&no\\
        dataset usage per epoch&$10\%$&$1\%$&$100\%$\\
        loss weights $(\lambda_1,\lambda_2,\lambda_3)$&$(1,0,0)$&$(100,2,1)$&$(100,2,1)$\\
        number of epochs&$100$&$50$&$50$\\
        warm-up epochs&$10$&$10$&$10$\\
        batch size: (GPUs $\times$ samples)&$4\times32$&$4\times32$&$4\times32$\\
        base learning rate&$\num{1e-3}$&$\num{1e-3}$&$\num{1e-4}$\\
        \bottomrule
    \end{NiceTabular}
    \end{adjustbox}
    \label{tab:curriculum}
\end{table}

%% file: tables/ast_performance.tex
\begin{table}
    \centering
    \vspace{-6pt}
    \caption{Performance obtained with different audio encoders on the binaural dataset described in \cite{zheng2024bat}. Numbers are percentages and the best performance in every column of each block is in bold.}
    \begin{adjustbox}{max width=\columnwidth}
    \begin{NiceTabular}{l cccc}
        \toprule
        &\multicolumn{4}{c}{performance metrics}\\
        \cmidrule(lr){2-5}
        audio encoder & \acs{map} ($\uparrow$) & \acs{er20} ($\downarrow$)  & \acs{mae} ($\downarrow$) & \acs{der} ($\downarrow$)\\
        \midrule
        SpatialAST \cite{zheng2024bat}: official checkpoint (2 stages) & $49.90$ & $24.43$ & $17.87$ & $32.50$\\
        \acs{dspast}: stage 1 & $53.05$ & $98.56$ & $95.57$ & $97.58$\\
        \acs{dspast}: stage 2 & $52.64$ & $20.31$ & \pmb{$14.44$} & $28.35$\\
        \acs{dspast}: stage 3 & \pmb{$54.53$} & \pmb{$20.28$} & \pmb{$14.44$} & \pmb{$28.03$}\\
        \midrule
        \acs{dspast}: original features & $54.76$ & $23.76$ & $17.76$ & \pmb{$26.45$}\\
        \acs{dspast}: original features, no feature attention & $52.88$ & $23.87$ & $17.55$ & $27.81$\\
        \acs{dspast}: only \acs{sed} loss for stage 2 and stage 3 & \pmb{$55.04$} & $97.19$ & $84.24$ & $68.18$\\
        \acs{dspast}: only \acs{doae} loss for stage 2 and stage 3 & $1.78$ & \pmb{$20.28$} & \pmb{$14.25$} & $81.98$\\
        \acs{dspast}: only \acs{dp} loss for stage 2 and stage 3 & $4.57$ & $97.76$ & $85.60$ & $27.28$\\
        \bottomrule
    \end{NiceTabular}
    \end{adjustbox}
    \label{tab:ast-performance}
\end{table}

%% file: tables/bat_performance.tex
\begin{table*}
    \centering
    \vspace{-6pt}
    \caption{Spatial perception and reasoning performance obtained with BAT based on different audio encoders for various question types of SpatialSoundQA \cite{zheng2024bat}. Numbers are percentages and the best performance in each column is in bold.}
    \begin{threeparttable}
    \begin{adjustbox}{max width=\textwidth}
    \begin{NiceTabular}{l ccccccccc}
        \toprule
        & \multicolumn{2}{c}{\acl{sed}: \acs{map} ($\uparrow$)}&\multicolumn{2}{c}{\acl{doae}: \acs{acc} ($\uparrow$)}&\multicolumn{2}{c}{\acl{dp}: \acs{der} ($\downarrow$)}&\multicolumn{3}{c}{spatial reasoning (type E): \acs{ba} ($\uparrow$)}\\
        \cmidrule(lr){2-3}\cmidrule(lr){4-5}\cmidrule(lr){6-7}\cmidrule(lr){8-10}
        audio encoder & type A & type C  & type B & type D & type B & type D & direction & distance & average \\
        \midrule
        Random guessing \cite{zheng2024bat} & $0.61$ & $0.59$ & $12.57$ & $12.41$ & $67.33$ & $67.46$ & $50.00$ & $50.00$ & $50.00$ \\
        SpatialAST \cite{zheng2024bat}: single-stage BAT & $24.18$ & $7.95$ & $72.59$ & $34.80$ & $33.61$ & $53.40$ & $67.53$ & $80.56$ & $74.04$ \\
        SpatialAST \cite{zheng2024bat}: multi-stage BAT\tnote{a} & $24.50$\tnote{b} & $7.97$\tnote{b} & $72.73$ & $35.08$ & $34.10$ & $52.81$ & $69.81$ & $80.29$ & $75.05$\\
        \midrule
        \acs{dspast}: single-stage BAT & $\pmb{27.15}$ & $\pmb{10.62}$ & $\pmb{78.84}$ & $\pmb{38.69}$ & $\pmb{28.41}$ & $\pmb{47.89}$ & $\pmb{72.52}$ & $\pmb{82.91}$ & $\pmb{77.71}$ \\
        \acs{dspast}: multi-stage BAT & $26.53$ & $9.22$ & $77.83$ & $38.25$ & $29.70$ & $50.95$ & $71.63$ & $81.21$ & $76.42$ \\
        \bottomrule
    \end{NiceTabular}
    \end{adjustbox}
    \begin{tablenotes}\footnotesize
    \item [a] The performance is based on the BAT implementation published by the authors (official checkpoint), which is worse than reported in \cite{zheng2024bat}.
    \item [b] These results are different from the official results, since we used different text embeddings that are free to use.
    \end{tablenotes}
    \end{threeparttable}
    \label{tab:bat-performance}
\end{table*}

%% file: figures/att_wts.tex
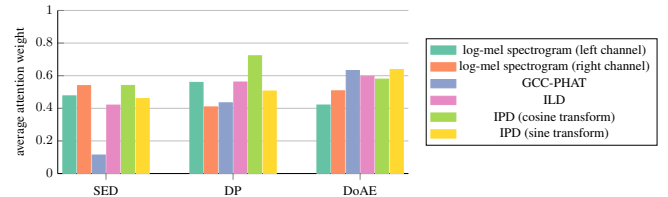
\begin{figure}[t]
    \centering
    \begin{adjustbox}{max width=\columnwidth}
    \begin{tikzpicture}
    \begin{axis}[
    axis y line*=left,
    axis x line*=bottom,
    ybar=1,
    area legend,
    ymin=0,
    ymax=1,
    xmin=0.85,
    xmax=3.15,
    enlarge x limits=0.1,
    point meta=explicit symbolic,
    legend style={at={(1.05,0.5)},anchor=west,legend columns=1},
    ylabel=average attention weight,
    bar width=9pt,
    height=5.5cm,
    width=10cm,
    xticklabels={\acs{sed}, \acs{dp}, \acs{doae}},
    xticklabel style={align=center},
    yticklabel style={align=center},
    xtick=data,
    typeset ticklabels with strut,
    xlabel near ticks,
    ylabel near ticks,
    nodes near coords,
    ymajorgrids,
    cycle list/Set2
    ]
    \addplot+[fill] coordinates {(1,0.47635007)(2,0.5589943)(3,0.42029357)};
    \addplot+[fill] coordinates {(1,0.5391766)(2,0.40858036)(3,0.50725496)};
    \addplot+[fill] coordinates {(1,0.11360918)(2,0.4337762)(3,0.63167536)};
    \addplot+[fill] coordinates {(1,0.41986382)(2,0.56126076)(3,0.5963022)};
    \addplot+[fill] coordinates {(1,0.5402806)(2,0.7221783)(3,0.57914495)};
    \addplot+[fill] coordinates {(1,0.4595191)(2,0.5059444)(3,0.6376595)};
    \legend{log-mel spectrogram (left channel), log-mel spectrogram (right channel), \acs{gcc}, \acs{ild}, \acs{ipd} (cosine transform), \acs{ipd} (sine transform)}
    \end{axis}
    \end{tikzpicture}
    \end{adjustbox}
    \caption{Average feature attention weights of \ac{dspast} on the test set.}
    \label{fig:att_wts}
\end{figure}